\documentclass{article}

    \usepackage[preprint]{neurips_2025}

\usepackage[utf8]{inputenc} % allow utf-8 input
\usepackage[T1]{fontenc}    % use 8-bit T1 fonts
\usepackage{hyperref}       % hyperlinks
\usepackage{url}            % simple URL typesetting
\usepackage{booktabs}       % professional-quality tables
\usepackage{amsfonts}       % blackboard math symbols
\usepackage{nicefrac}       % compact symbols for 1/2, etc.
\usepackage{microtype}      % microtypography
\usepackage{xcolor}         % colors
\usepackage[pdftex]{graphicx}
\usepackage{amsmath}
\usepackage{bm}
\usepackage{algorithm}
\usepackage{algorithmic}
\usepackage{multirow}
\usepackage{pifont}% http://ctan.org/pkg/pifont
\usepackage{changes}
\newcommand{\cmark}{\ding{51}}%
\newcommand{\xmark}{\ding{55}}%

\title{Connectivity determines the capability of \\ sparse neural network quantum states}

\author{%
  Brandon Barton$^{1,2}$,
  Juan Carrasquilla$^{1}$,
  Christopher Roth$^{2}$\thanks{These authors contribitued equally to this work.} ,
  Agnes Valenti$^{2*}$\\
$^{1}$Institute for Theoretical Physics, ETH Zürich, Zürich 8093, Switzerland\\
$^{2}$Center for Computational Quantum Physics, Flatiron Institute, New York, NY 10010, USA\\
\texttt{\{bbarton,jcarrasquilla\}@ethz.ch}, \texttt{\{croth,avalenti\}@flatironinstitue.org}
}

\begin{document}

\maketitle

\begin{abstract}
  The Lottery Ticket Hypothesis (LTH) posits that within overparametrized neural networks, there exist sparse subnetworks that are capable of matching the performance of the original model when trained in isolation from the original initialization. We extend this hypothesis to the unsupervised task of approximating the ground state of quantum many-body Hamiltonians, a problem equivalent to finding a neural-network compression of the lowest-lying eigenvector of an exponentially large matrix. Focusing on two representative quantum Hamiltonians, the transverse field Ising model (TFIM) and the toric code (TC), we demonstrate that sparse neural networks can reach accuracies comparable to their dense counterparts, even when pruned by more than an order of magnitude in parameter count. Crucially, and unlike the original LTH, we find that performance depends only on the structure of the sparse subnetwork, not on the specific initialization, when trained in isolation. Moreover, we identify universal scaling behavior that persists across network sizes and physical models, where the boundaries of scaling regions are determined by the underlying Hamiltonian. At the onset of high-error scaling, we observe signatures of a sparsity-induced quantum phase transition that is first-order in shallow networks. Finally, we demonstrate that pruning enhances interpretability by linking the structure of sparse subnetworks to the underlying physics of the Hamiltonian.
\end{abstract}

\section{Introduction}
\label{sec:introduction}
Over the past decades, pruning has emerged as popular approach to reduce the number of parameters of neural networks \cite{lecun1989optimal, hassibi1992second, reed1993pruning, han2015learning}. The goal of pruning is to speed up model inference and improve portability without compromising on accuracy. In practice, contemporary models can be pruned to $50 \%$ or even $90 \%$ sparsity while mostly retaining their original performance \cite{gordon2020compressing, gan2022playing, yeo2023can}. Pruning has also led to crucial insight on the inner workings of neural networks with many parameters: The lottery ticket hypothesis (LTH) states that sparse sub-networks (``winning tickets'') are contained within densely parametrized models that can match or exceed the performance of the original model when trained in isolation with proper initialization \cite{frankle2018lottery}. The LTH has been verified in subsequent experimental and theoretical studies \cite{frankle2019stabilizing, diffenderfer2021multi, savarese2020winning, chen2020long} as well as extended to different types of neural networks \cite{yu2019playing, chen2020lottery, chen2021unified, zhang2024graph}.

We here examine the LTH in a non-traditional use-setting: The task of approximating the ground state of a quantum many-body system with a neural network \cite{carleo2017solving, carrasquilla2017machine}. Our task can be abstracted to the problem of learning a mapping from discrete data (spin-configurations $(\sigma_1, ... \sigma_N)$, where $\sigma_i \in \{+1, -1\}$ and $N$ is the number of spin degrees of freedom) to a set of real numbers, called {\it amplitudes}, which specifies the ground state eigenvector of a many-body Hamiltonian. 

The object we represent is profoundly different from that of typical machine learning tasks such as image recognition \cite{krizhevsky2017imagenet}, or language modeling \cite{naveed2023comprehensive}. There, a neural network is asked to represent extremely sparse objects, such as the subset of all RGB images that contain a cat, or the subset of random vocabulary strings that form coherent sentences. In contrast, a quantum many-body wavefunction is in general a fully dense representation, with probability amplitudes for different configurations spanning many orders of magnitudes. In order to capture the correct physics, these amplitudes must be encoded with high precision which typically necessitates the use of a more performant training algorithm than stochastic gradient descent. A common choice is natural gradient descent, which accounts for the curvature of the neural-network function.

A hidden benefit of considering physical systems is that we can readily apply physics-based analysis tools to better study the pruning dynamics. Since our neural network directly encodes a physical phase, we can use the language and mathematics of quantum phase transitions as a tool to understand the different sparse representations that we find with pruning. 

In order to look for robust principles, we vary both the neural network architecture and the physical Hamiltonian. Throughout this work, we consider two distinct physical models: the transverse field Ising model for different values of transverse field and the topological toric code model \cite{kitaev2003fault}. We test the LTH on these cases and analyze the behavior of the pruned networks in terms of the physics of the ground state wavefunction. 

Surprisingly, we find that an even stronger version of the LTH holds for the ground state search problem on the Hamiltonians we consider in this work. By using a learned sparse structure with $5-20 \%$ weights of the dense model, we are able to match the performance of a fully dense model using a random initialization of weights. We find that the optimal sparsity depends on the physics of the considered quantum many-body model. Additionally, we see that starting from the weights from the ``lottery ticket initialization'' offers negligible benefit over random weights. Crucially, the {\it connectivity} of the sparse masks by itself constitutes a valid lottery ticket independent of its initialization  (see Section (~\ref{subsec:LTH})). 

In Section~(\ref{subsec:sparse_scaling_behavior}), we proceed to examine the scaling behavior of shallow FFNNs as a function of pruning and identify scaling regions that are universal with respect to varying initial network size, in accordance with \cite{pmlr-v139-rosenfeld21a}: At both low and high sparsities, we find plateaus where the accuracy is fairly independent of the number of remaining parameters. Given that these plateaus occur at very different variational energies, they are likely representing different phases of matter. We gain further insights into the nature of these plateaus, and what happens in between, using tools to characterize phases of quantum matter. In particular, we show in Section~(\ref{sec:phase_transitions}) that the low-and high-error plateau correspond to physical phases of matter, connected by at least one (first-order) phase transition and possibly a third intermediate phase.

In Section~(\ref{subsec:interpretability}), we then proceed to relate the structure and weights of the pruned sub-networks to the physics of the underlying model. We show that our pruning procedure discovers an efficient solution to the toric code problem with an error that can be made asymptotically small.

\paragraph{Contributions} Our contributions are as follows:
\begin{itemize}
\item We show that neural network quantum many-body wavefunctions have a different form of the LTH. Here, the lottery tickets are sparse subnetworks, and the initial weights are not relevant to the final performance. This contrasts with the LTH in other problem domains \cite{frankle2018lottery, chen2020lottery}, where a lottery ticket consists of a sparse subnetwork initialized with particular parameter values.

\item We demonstrate that neural network wavefunctions have distinct scaling regions across several quantum many-problems, consisting of high-and low error plateaus that bookend an approximately power law region. Changing the number of parameters only affects the performance of the models in the low-error plateau.
\item We introduce quantum physics-based tools for interpreting pruning behavior in terms of phase transitions. We use this to identify a first-order phase transition as a function of sparsity for shallow FFNNs trained on the TFIM. 
\item We find an asympotically exact solution to the toric code through pruning, which we elucidate in detail. When the toric code is pruned past the minimum sparsity for this solution, it undergoes a phase transition.
\end{itemize}

\section{Related work}

\paragraph{Pruning neural-network quantum states}
In the context of neural-network quantum states, the effect of pruning has been studied for the task of reconstructing a known quantum many-body wave-function from a 1D Hamiltonian \cite{golubeva2022pruning, sehayek2019learnability}, which is a supervised learning task. In contrast, we study an unsupervised learning task of finding the minimum energy variational wavefunction of 2D Hamiltonians. Our work also interprets the structure of the learned sparse representations in light of the physical Hamiltonian.

\paragraph{Scaling laws}
This work is inspired by previous studies on predictable scaling laws \cite{bahri2024explaining} of sparse neural networks during iterative pruning \cite{pmlr-v139-rosenfeld21a}. We test these methods on a non-traditional task of learning a quantum ground state. For neural quantum states, it is important to understand how the number of parameters necessary to achieve a desired relative error scales with system size. While this relation has been studied for dense models \cite{chen2024empowering, passetti2023can, denis2023comment}, here we examine scaling behavior through the lens of {\it sparse} networks by pruning initially dense models and analyzing the resulting error scaling.

\paragraph{Structure of sparse subnetworks}
Following the LTH, a stronger hypothesis \cite{ramanujan2020s} was proven \cite{malach2020proving, pensia2020optimal, orseau2020logarithmic, burkholz2021existence, cunha2022proving, ferbach2022general}, showing that high-performing sparse subnetworks already exist at initialization (``strong lottery tickets") for a variety of neural network architectures. Traditionally, strong lottery tickets are therefore characterized by both a sparse subnetwork and a set of initial parameter values. In support of this, it has also been shown in the original LTH that changing the initialization of lottery tickets results in a performance decrease \cite{frankle2018lottery}. Our evidence points to the contrary for the case of neural-network quantum states, as the lottery ticket performance only depends on its structure, not its parameter initialization. Other related studies have shown that parameter sign configurations play an important role in obtaining winning tickets \cite{gadhikar2024masks, zhou2019deconstructing, oh2025find}. In this work, we are able to find sparse subnetwork representations that correspond to asymptotically exact solutions of the problem at hand in both absolute values of the weights as well as their sign structure.

\paragraph{Phase transitions in neural networks}
Several works have used tools from statistical mechanics in order to analyze neural network representations. One approach understands phase transitions as a function of training time, by considering the neural network as a classical Hamiltonian with individual neurons as spin variables, and the weights as time-dependent coupling strengths as a function of training under stochastic gradient descent \cite{barney2024neural, winer2025deep}. Another perspective characterizes phase transitions in a neural network's generalization capabilities as a function of training samples by employing replica methods from statistical mechanics \cite{sompolinsky1990learning, gyorgyi1990first, barbier2019optimal, schwarze1993learning, maillard2020phase, barra2017phase, cui2024phase}. 
Our work makes a direct connection to phase transitions in quantum many-body physics -- we characterize phases directly by computing order parameters on a many-body wavefunction parameterized by a neural network. We find that these order parameters can change as a function of sparsity.

\section{Experimental set-up}
\label{sec:exp_setup}

\subsection{Neural network representations of quantum states}
\label{subsec:NQS}

Throughout this work, we consider systems with $N$ spin-$1/2$ degrees of freedom.
We aim to solve the ground-state problem. In particular, we are given a Hamiltonian $\hat{H}$ for which we want to find the lowest-lying eigenstate, i.e.
\begin{align}
\hat{H}|\Psi\rangle_G = E_G |\Psi\rangle_G,
\end{align}
where $E_G$ is the system's ground state energy.
In general, the dimensionality of this problem renders an exact solution unfeasible as the Hilbert space size scales as $2^N$.

Neural network quantum states have emerged as a method to circumvent this exponential scaling. Concretely, we parametrize the many-body quantum state by expanding it in a basis of spin-configurations
\begin{equation}
    \vert \Psi^{NN}_{\theta} \rangle = \sum_{\bm{\sigma}} \Psi^{NN}_{\theta}(\bm{\sigma}) \vert \bm{\sigma} \rangle.
\end{equation}
Here, the wave-function amplitudes $\Psi^{NN}_{\theta}(\bm{\sigma})$ are parametrized with the help of a neural network: The output value of the network is chosen to represent $\log(\Psi_{\theta}^{NN}(\bm{\sigma}))$ when given as input a spin-configuration $\bm{\sigma} = (\sigma_1, ... \sigma_N)$, $\sigma_i \in \{ \pm 1 \}$. The weights and biases of the network form the set of parameters $\theta$. Throughout this work, we only consider wave functions that are known to be real and positive. Our experiments are performed on three different network architectures: shallow feed-forward neural networks (FFNNs), shallow convolutional neural networks (CNNs) and few-layer residual convolutional neural networks (ResCNNs). More details on the employed neural networks and hyperparameters can be found in the Appendix. 

The neural network is trained to approximate the ground state by making use of the variational principle and minimizing the cost function
\begin{equation} \label{eq:varen}
C(\theta) = \langle \hat{H} \rangle_{\theta} \geq E_{\mathrm{G}}.
\end{equation}
Here, $\langle \cdot \rangle_{\theta}$ denotes the expectation value taken with respect to the wave-function $|\Psi^{NN}_{\theta}\rangle$. All expectation values are computed via Monte Carlo sampling.
 We minimize this cost function via natural gradient descent, in the context of the ground-state search also termed ``stochastic reconfiguration'' (SR) \cite{sorella1998green}. For more details on sampling and optimization, see Appendix. The simulations are carried out with the help of the package {\it netket} \cite{carleo2019netket, vicentini2022netket}, based on  JAX \cite{bradbury2018jax} and FLAX \cite{flax2020github}.

Given a so-obtained ground-state approximation $|\Psi^{NN}_{\theta}\rangle$, we probe its accuracy using the relative error 
\begin{equation}\label{eq:rel_err}
    \epsilon_{\mathrm{rel}}(E) = \left| \frac{ E_{\mathrm{exact}} - \langle \hat{H} \rangle_{\theta}}{ E_{\mathrm{exact}}} \right|,
\end{equation}
where the reference exact ground-state energy $E_{\mathrm{exact}}$ is computed either via exact diagonalization on a small system, the density-matrix-renormalization group (DMRG) or the lowest energy that was obtained with our neural network quantum state. In some cases throughout this work, we choose to report the absolute error per spin $\Delta E = \vert E_{\mathrm{exact}} - \langle \hat{H} \rangle_{\theta} \vert / N$, when comparing errors in different quantum phases or across multiple system sizes. All DMRG computations are carried out using the package ITensor \cite{itensor}. The implementation details can be found in the Appendix.

\subsection{Physical models}
\label{subsec:physical_models}

We study the effect of pruning on the problem of the ground-state search for two different physical models which we introduce below: The transverse-field Ising model and the toric code model. We mainly focus on the transverse-field Ising model due to its versatility, and utilize the toric code model because of its exact solubility to interpret obtained sparse subnetworks.

\paragraph{Transverse-field Ising model}
We consider the transverse-field Ising model in two dimensions, with $N = L^2$ spin $1/2$-degrees of freedom on a $L \times L$ lattice. The system is described by the Hamiltonian 
\begin{equation}
  \hat{H}_{\mathrm{TFIM}} = -\sum_{\langle i, j \rangle} \hat{\sigma}_{i}^z \hat{\sigma}_{j}^z - \kappa \sum_{i} \hat{\sigma}_i^x,
  \label{eq:tfim_ham}
\end{equation}
with the Pauli matrices $\hat{\sigma}^{x/z}$.
The interplay between spin coupling of nearest neighboring spins (denoted by $\langle i,j \rangle$) and the transverse magnetic field $\kappa > 0$ determines the quantum phase of the model. In particular, the Ising model exhibits a quantum phase transition as a function of magnetic field $\kappa$:
For small magnetic field, the spin coupling term dominates and the ground state is ferromagnetic. In the case of a strong transverse field $\kappa$, spins are aligned in form of a paramagnet in the $+x$ direction. The quantum phase transition is estimated to occur at $\kappa_c \approx 3.04438(2)$ \cite{PhysRevE.66.066110}. At the position of the phase transition, the system is at criticality and spin-spin correlation functions decay with power-law.

\paragraph{Toric code model}
As a use-case of a model with an exact solution and topological order, we consider the toric code model \cite{kitaev2003fault}. The toric code model is defined on a $L \times L$ lattice with periodic boundary conditions and $N=2L^2$ spin-$1/2$ degrees of freedom on the edges. The Hamiltonian is given by
\begin{equation}
    \hat{H}_{\mathrm{TC}} = - \sum_{v} \hat{A}_v - \sum_{p} \hat{B}_p,
    \label{eq:tc_ham}
\end{equation}
where $\hat{A}_v = \prod_{i \in v}\hat{\sigma}^x_i$ and $\hat{B}_p = \prod_{i \in p} \hat{\sigma}^z_i$ act on the plaquettes and vertices on the given lattice. The ground-state manifold is four-fold degenerate and consists of eigenstates of the mutually commuting $\hat{A}_v$ and $\hat{B}_p$ with eigenvalue $+1$.  
The toric code serves as instructive example to test interpretability of neural-network quantum states, as several exact constructions of the toric code ground states with the help of neural networks have been brought forward in previous work \cite{deng2017machine, carrasquilla2017machine,valenti2022correlation, chen2025representing}.

\subsection{Sparse training strategies}
\label{subsec:sparse_training_strat}

The employed pruning strategy throughout this work is iterative magnitude pruning with weight rewinding (IMP-WR). We comment on this choice and compare it to other pruning strategies in the Appendix. We detail the pruning algorithm and describe isolated training strategies that are carried out using the sparse masks obtained via IMP-WR below.

\paragraph{Iterative magnitude pruning with weight rewinding}
Concretely, the IMP-WR procedure begins with training a randomly initialized dense neural network (pre-training phase). The optimized parameters $\theta_{\mathrm{wr}}$ from this pre-training phase serve as \emph{rewinding point}: Once this rewinding point is obtained, the IP-WR procedure proceeds in an iterative fashion. In each pruning iteration, the network's remaining non-zero weights are initialized in the rewinding point and then trained. After training, a fraction $p_r$ of the weights with smallest magnitude removed in an unstructured fashion.  We use a pruning ratio of $p_r =0.12$ for FFNNs and ResCNNs, and a smaller pruning ratio of $p_r=0.05$ for CNNs due to their initially lower parameter count. This process is repeated for a total of $I$ pruning iterations. 
The number of remaining parameters $n$ of the neural network at a particular iteration $i$ is given by
\begin{equation}\label{eq:density}
    n := (1-p_{\mathrm{r}})^{i} n_{\mathrm{init}},
\end{equation}
where $n_{\mathrm{init}}$ is the number of parameters in the initial network. Throughout this work, we report the number of remaining parameters $n$, and investigate the ability of the NQS to approximate the ground state of physical models as the neural network becomes increasingly sparse. 
Further details on the pruning procedure are provided in the Appendix.

\paragraph{Isolated training variants}
To test the lottery ticket hypothesis, we perform different variants of isolated training from sparse initialized networks (``tickets'' $T$). In particular, we denote the initialization of such a sparse subnetwork as a combination of paramaters $\theta$ and a pruning mask $m \in \{0,1\}^{n_{\mathrm{init}}}$. Then, the parameters of the ticket $T(\theta,m)$ are given by element-wise product between $\theta$ and the pruning mask $m$. We then proceed to train $\theta$ in isolation, while keeping the pruning mask $m$ fixed. In particular, we define three distinct such tickets:
\begin{itemize}
    \item $T(\theta_{\mathrm{init}}, m_{\mathrm{imp}})$: Initialized to initial IMP-WR parameters $\theta_{\mathrm{init}}$ with masks $m_{\mathrm{imp}}$, obtained from IMP-WR at a selected pruning iteration.
    \item $T(\theta_{\mathrm{rand}}, m_{\mathrm{imp}})$: Initialized to random parameters $\theta_{\mathrm{rand}}$ with masks $m_{\mathrm{imp}}$, obtained from IMP-WR at a selected pruning iteration.
    \item $T(\theta_{\mathrm{init}}, m_{\mathrm{rand}})$: Initialized to initial IMP-WR parameters ${\theta}_{\mathrm{init}}$ with random sparse masks $m_{\mathrm{rand}}$.
\end{itemize}
The $T({\theta}_{\mathrm{init}}, m_{\mathrm{imp}})$ variant tests the original LTH, while $T({\theta}_{\mathrm{rand}}, m_{\mathrm{imp}})$ tests the viability of using the same mask on a new parameter initialization. Finally, $T({\theta}_{\mathrm{init}}, m_{\mathrm{rand}})$ is a random sparse initialization of the pruning mask, which serves as a benchmark to the other two isolated training variants. The isolated training variants are tested with masks derived from single trajectories of iterative magnitude pruning, and we report the statistics over several random initializations in the Appendix.

\section{Results}
\label{sec:results}

\subsection{Testing the lottery ticket hypothesis}
\label{subsec:LTH}

\begin{figure}[h]
    \centering
    \includegraphics[width=\linewidth]{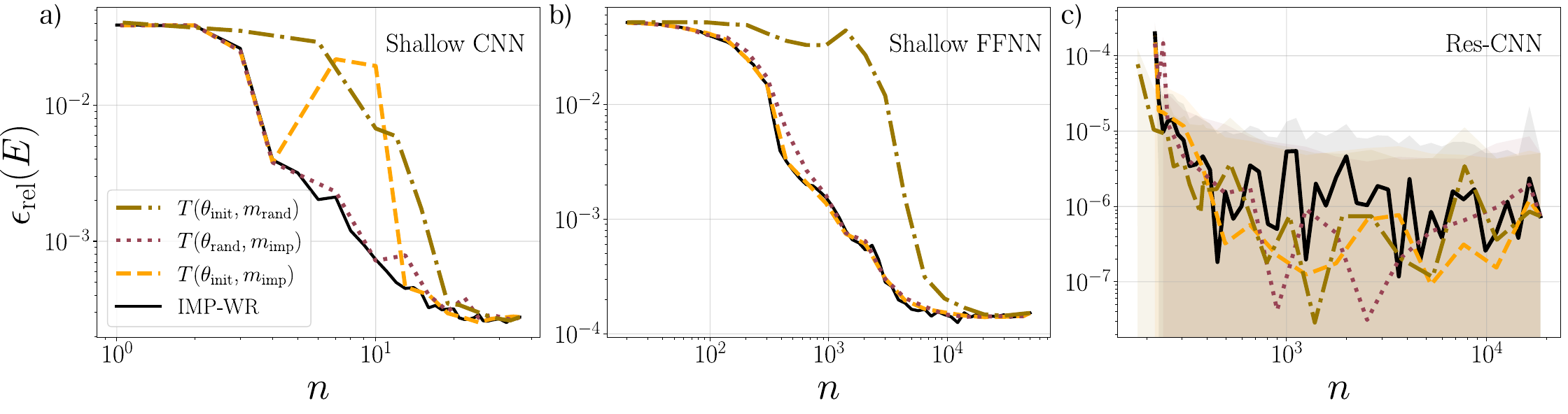}
    \caption{Test of the LTH at the critical point $\kappa = \kappa_c$ of the TFIM on the $N=10\times10$ lattice, for various neural network architectures: (a) shallow CNN, (b) shallow FFNN, and (c) Res-CNN (shaded regions are statistical sampling errors). The relative error is computed with respect to the average energy of three best-performing dense ResCNNs by smallest variance (see Appendix).}
    \label{fig: LTH testing}
\end{figure}

We test the validity of the LTH for the case of neural-network quantum states. For this purpose, we follow the sparse training strategies outlined in Section~\ref{subsec:sparse_training_strat}. In particular, we obtain the sparse masks $m_{\mathrm{imp}}$ used for our lottery ticket tests by applying IMP-WR on a neural network initialized with parameters $\theta_{\mathrm{init}}$. We then proceed with the isolated training of $T(\theta_{\mathrm{init}},m_{\mathrm{imp}})$, $T(\theta_{\mathrm{rand}},m_{\mathrm{imp}})$ and $T(\theta_{\mathrm{init}},m_{\mathrm{rand}})$. The performances using various different model architectures are shown in Fig.~\ref{fig: LTH testing} for the case of the TFIM at criticality ($\kappa = \kappa_c$) on a $N=10\times 10$ lattice.

The original LTH \cite{frankle2018lottery} asserts that it is the combination of masks {\it and} initial parameter values that forms a winning ticket. Thus we would expect training $T(\theta_{\mathrm{init}}, m_{\mathrm{imp}})$ in isolation to greatly outperform the other ticket initialization choices. While we find that winning ticket networks can replicate the performance of the dense network (up to a critical sparsity), we also observe that the same sparse masks perform equally well when parameters are randomly initialized. On the other hand, we find that sparse networks based on masks with randomly chosen connectivity perform significantly more poorly; the performance begins to degrade with a small amount of pruning, and they reach high error much faster (see Fig.~ \ref{fig: LTH testing}).

In the case of the ResCNN architecture,  the performance is fairly independent of both the initial parameters and the learned masks. Most likely this behavior has it roots in the starting point: the initial kernels are of size $3\times 3$. This very localized structure already provides an extremely strong starting point, as we can encode the ground state nearly exactly with O($10^3$) parameters.

From these experiments we conclude that a lottery ticket encoding of a quantum wavefunction is simply a sparse connectivity mapping. In Section~\ref{subsec:interpretability} we elucidate this observation by showing that the connective structure of our pruned models can be understood in terms of the Hamiltonian that we simulate.

\subsection{Sparse scaling behavior}
\label{subsec:sparse_scaling_behavior}

We analyze the behavior of the neural network and the corresponding quantum wave-function as a function of pruning more in detail. To this end, we focus on the simplest use-case, the shallow FFNN of depth $d=1$. For more details on the CNN and ResCNN scaling, see Appendix.

\paragraph{Universal scaling behavior}

\begin{figure}[h]
    \centering
    \includegraphics[width=\linewidth]{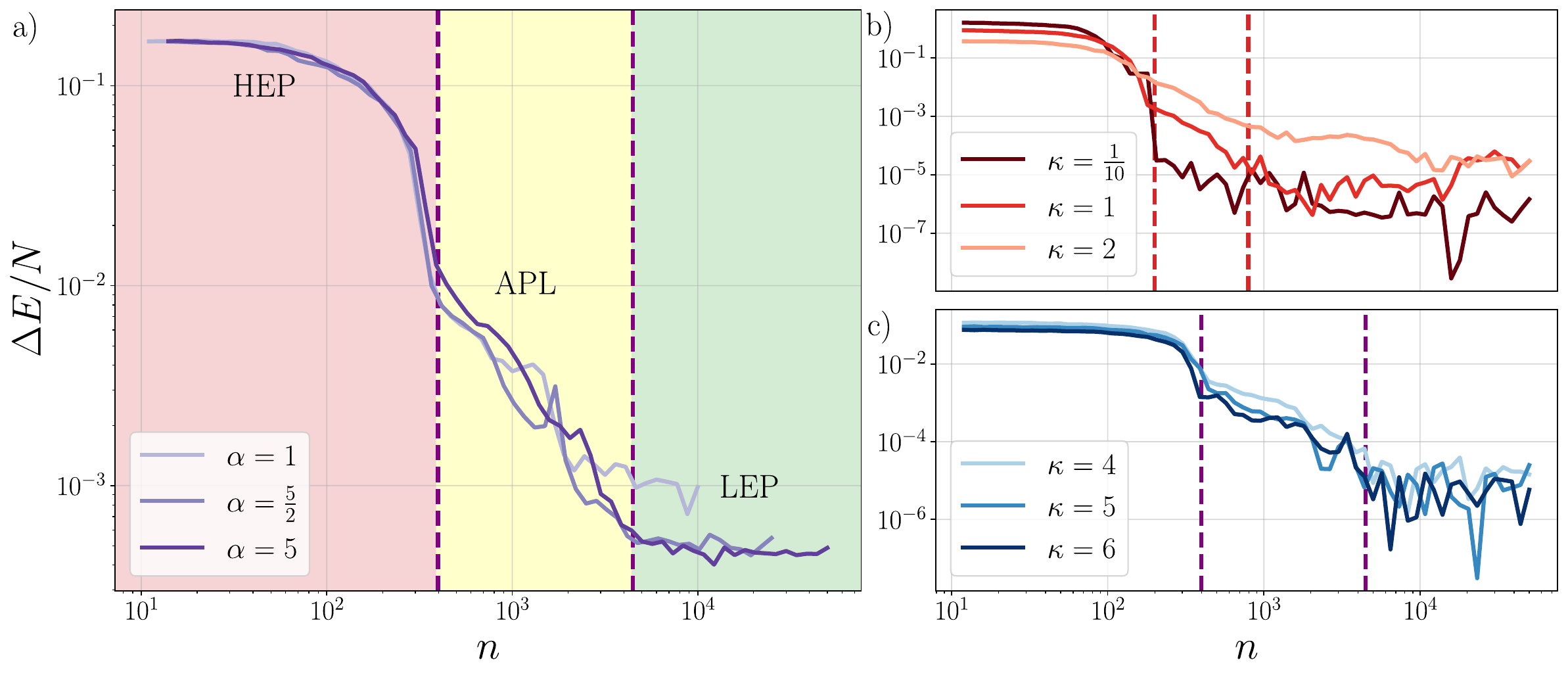}
    \caption{Plotted are regions of error scaling within IMP-WR for the TFIM on the $N=10 \times 10$ lattice using FFNNs with depth $d=1$. The subplots correspond to simulations at (a) fixed $\kappa$ (at the critical point $\kappa=\kappa_c$) with varying network widths $w=\alpha N$  (b) varying $\kappa<\kappa_c$ within the ferromagnetic phase and fixed width $w = 5N$ (b) varying $\kappa>\kappa_c$ within the paramagnetic phase and fixed width $w = 5N$. We qualitatively highlight the scaling regions via colors in (a) and dashed lines in (b) and (c).}
    \label{fig: regions of scaling}
\end{figure}

In accordance with \cite{pmlr-v139-rosenfeld21a}, the scaling behavior of the pruned FFNN error shown in Fig.~\ref{fig: LTH testing} suggests three scaling regimes: (i) a low-error plateau (LEP) at high densities $n$ (ii) a high-error plateau (HEP) at low densities and (iii) an approximate power-law region (APL) connecting the two plateaus. 
We find the error scaling in the (HEP) and the (APL) to be {\it universal} with respect to the width of the initial network; Fig.~\ref{fig: regions of scaling} demonstrates evidence that the error of all shallow FFNNs of different width coalesce onto one another. However, the LEP at large number of remaining parameters shows a slight dependence on the initial number of network parameters: These denser networks are qualitatively different than the very sparse networks in the APL and HEP, involving a more complex structure that is broken within the APL. We will further elaborate on this question in Section~\ref{sec:phase_transitions}.

\paragraph{Universal scaling behavior within different phases of the TFIM}
Up to here, we have considered pruning of neural-network quantum states that approximate the ground state of the TFIM at criticality. This corresponds to setting the field strength $\kappa$ in the Hamiltonian to the critical value $\kappa=\kappa_c$. 
We now examine the behavior of the pruned FFNN for the ferromagnet ($0\leq \kappa < \kappa_c$) and the paramagnet ($\kappa > \kappa_c$). We find that each phase exhibits the same qualitative scaling behavior, i.e. showcases the three regions LEP, APL and HEP (see Fig.~\ref{fig: regions of scaling}). However, the boundaries of these regions are unique to each phase. In particular, the onset to the HEP occurs at lower sparsity in the case of the ferromagnet in comparison to the paramagnet and the critical state. A possible explanation lies in the sparsity of the true ground state eigenvector: while the paramagnet and critical phases have non-zero amplitudes for many different $\sigma_z$ magnetizations, the ferromagnet only has significant amplitudes when almost all of the spins are aligned. Representing this very sparse state with a neural network only requires very few nonzero weights (see Appendix). 

\subsection{Sparsity-induced phase transition}
\label{sec:phase_transitions}

So far, we have examined the generic neural-network scaling regions under pruning based on the behavior of the relative error. Now, we aim to get a better understanding of the pruning {\it dynamics}: For this purpose, we fix $\kappa=\kappa_c$ and study the evolution of the neural-network {\it quantum wave-function} as we vary the network sparsity. We are thereby able to link pruning dynamics of the neural network representation to physical quantum phases.

Fundamental properties in quantum many-body physics such as phases of matter are only well-defined in the thermodynamic limit, i.e. in the limit of an infinite number of (here spin 1/2)-degrees of freedom. In a machine-learning language, this translates into an input layer of infinite size.
We here define the thermodynamic limit by sending $n,N\to \infty$ (where $N$ is the number of spins) while at the same time keeping the ratio $\rho=n/N$ fixed.

\begin{figure}[h]
    \centering
    \includegraphics[width=\linewidth]{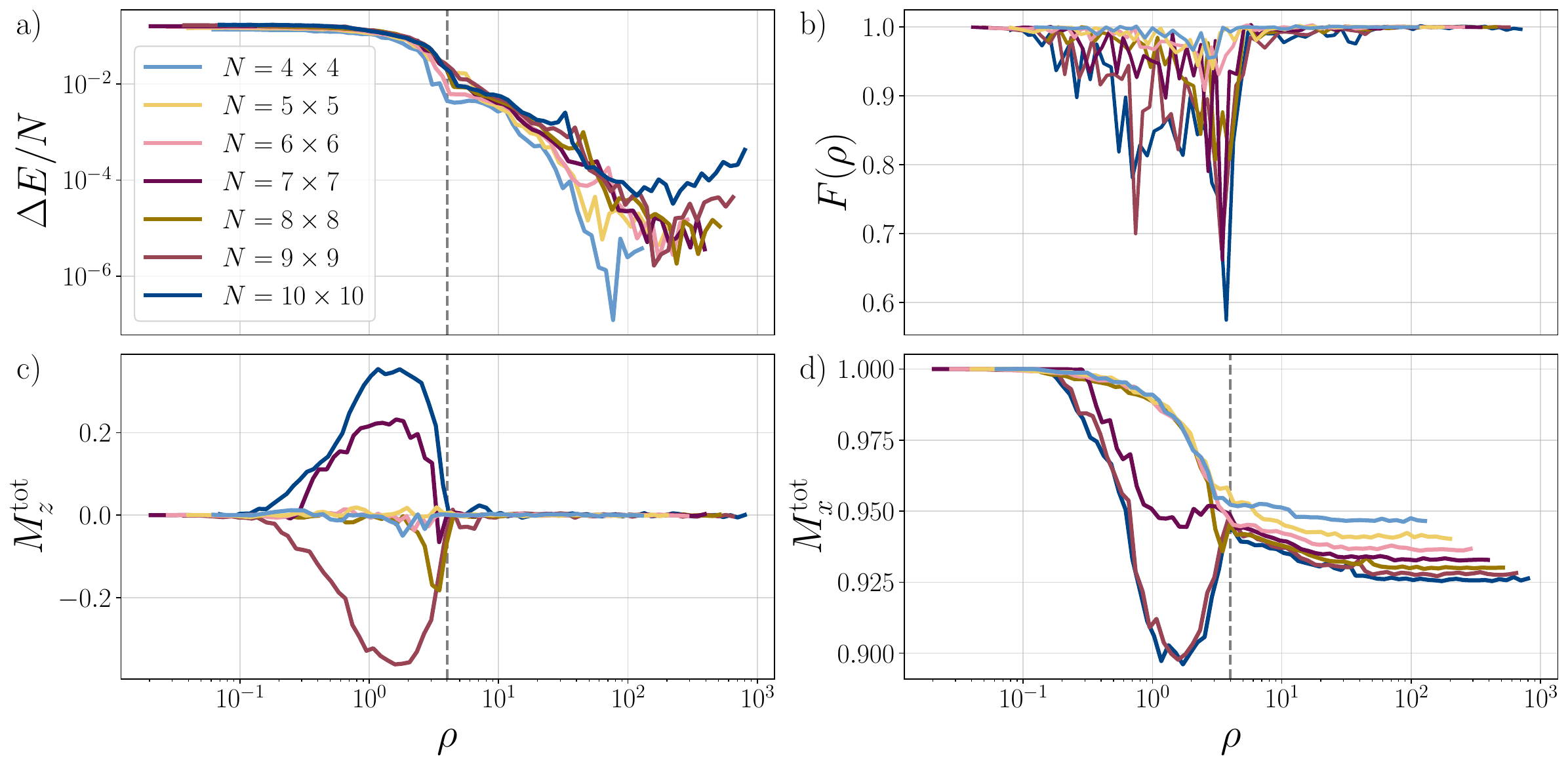}
    \caption{Evidence of a sparsity-induced quantum phase transition in shallow depth $d=1$ FFNNs. This data corresponds to finite size scaling at the quantum critical point $\kappa = \kappa_c$ of the TFIM for $N=4\times 4$ through $N=10 \times 10$ spins. We plot the (a) relative error per spin versus the number of parameters per spin (b) fidelity between neighboring pruned wave functions, (c,d) the total magnetization in the $x$ and $z$ directions. The phase transition is marked via a dashed line in  (a,c,d).}
    \label{fig: phase transition}
\end{figure}

We approach the thermodynamic limit via a finite-size scaling analysis: We repeat the pruning experiment for varying lattice (input layer) sizes. The results are depicted in Fig.~\ref{fig: phase transition}.
Crucially, when plotted as a function of the number of parameters per spin $\rho$, the error scaling curves at different lattice sizes coalesce. Thus, the observed features (i.e. the 3 scaling regimes) are well-defined in the thermodynamic limit and not an artifact of finite-size effects. Additionally, the transitions in between the regions occur at a scale-invariant number of parameters per spin. It is thus a natural question to ask, to what extent the scaling regions directly correspond to quantum phases of matter. 

For this purpose, we can think of $\rho$ as a phase-transition driving parameter, in analogy with the transverse field $\kappa$ of the Ising model. Just as we can study the physics of the Ising model as a function of $\kappa$ by computing correlation functions and order parameters, we can use the exact same toolkit to study $\Psi^{NN}(\rho)$.

What do we know about the quantum phase of $\Psi^{NN}(\rho)$ a priori? When $\rho$ is very large and the neural-network approximation is very good the quantum phase of $\Psi^{NN}(\rho)$ will correspond to the same phase as the targeted state $\Psi^{exact}_{\kappa=\kappa_c}$, i.e. the exact ground state of the TFIM at $\kappa=\kappa_c$. We expect this to be true in the LEP. However, $\Psi^{NN}(\rho)$ may undergo a phase transition to a phase that is easier to represent as $\rho$ is decreased. We find evidence of this behavior in a kink of the network error at the transition between the APL and the HEP. This signifies a discontinuity in  $\frac{d E}{d \rho}$, which is a typical signature of a (first-order) phase transition.

We sustain this claim by considering a more sensitive measure of quantum phase transitions, the {\it fidelity} $F(\rho)=\langle \Psi^{NN}(\rho) \Psi^{NN}(\rho + \epsilon) \rangle$, where $\epsilon$ is small (approaching zero). It measures the overlap between two quantum states states at adjacent points along the pruning trajectory. At the position of a phase transition, the two adjacent states are in two different phases and thus one shall expect the fidelity to approach zero. We find that the fidelity drops sharply at the transition between the APL to HEP scaling regions, providing further evidence for a phase transition. The transition we observe for the TFIM $(\kappa=\kappa_c)$ occurs at a critical parameter count around $\rho_c \approx 3 - 5$.

At the same critical number of parameters $\rho_c$ where the fidelity drops, there is a sudden change in the nature of the state, observed in the total $x$ and $z$ magnetization shown in \ref{fig: phase transition}. 
Here, the magnetization is given by $M_z^{\rm tot}=\langle \sum_i \sigma_i^z \rangle_{\theta}/N$ (and similarly for $M_x^{\rm tot}$). This sharp feature further points to the nature of the transition being first-order. We can understand this transition as a transition between a $\mathbb{Z}_2$-symmetric state (which corresponds to the symmetry of the exact ground state in a finite system) to a symmetry-broken state. No apparent phase transition occurs between the LEP and ALP regions, but the behavior within HEP itself changes drastically: The $z$-magnetization drops to zero as sparsity is increased. It is unclear if this is an adiabatic transition between phases or evidence for a second,  possibly continuous, phase transition.

\subsection{Interpretable sparse neural network solution to the toric code}
\label{subsec:interpretability}

\begin{figure}[h]
    \centering
    \includegraphics[width=\linewidth]{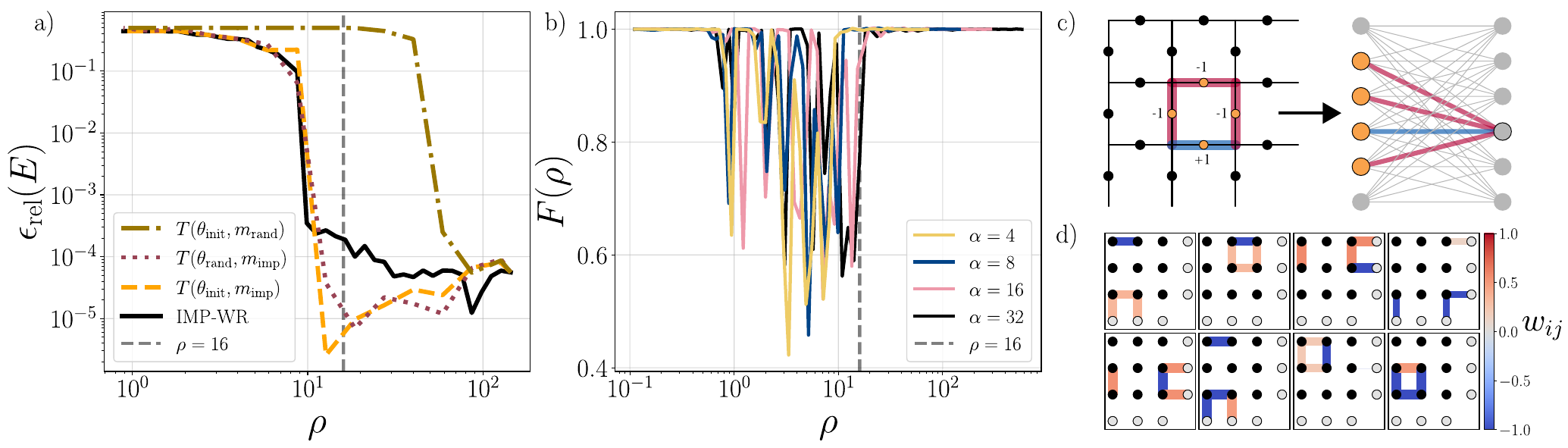}
    \caption{(a) LTH test for the toric code with $N=18$ spins using a depth $d=1$ FFNN of width $w=8 N$. (b) Fidelity between neighboring pruned NQS with dashed line at critical number of parameters $\rho_c = 16$.  (c) Schematic of a single hidden neuron connected to a single plaquette. (d) Normalized weights and their sign configuration from eight single hidden neurons in a pruned FFNN.}
    \label{fig: toric code}
\end{figure}

We find that we can naturally obtain an efficient solution to the toric code by pruning a single-layer FFNN. In Fig.~\ref{fig: toric code}, the $T(\theta_{\mathrm{init}}, m_{\mathrm{imp}})$ and $T(\theta_{\mathrm{rand}}, m_{\mathrm{imp}})$ isolated training tests outperform the iteratively pruned network by around two orders of magnitude in accuracy, while the 
$T(\theta_{\mathrm{init}}, m_{\mathrm{rand}})$ benchmark reaches a high-error plateau at a much higher relative parameter count. Aligning closely to our conclusions for the TFIM, these results demonstrate that the structure of these sparse subnetworks contain a highly efficient and accurate representation of the ground state of the toric code.

Similarly to the case of the TFIM, we find evidence for a phase transition in the pruning dynamics of the toric code: the fidelity $F$ between  neural networks of slightly different sparsity drops rapidly at a critical parameter density  of around $\rho_c =16$, as shown in Fig.~\ref{fig: toric code}. This critical density corresponds to 16 weights per spin, or 32 weights per plaquette. We can understand this density as an asymptotically exact solution to the toric code, where the error decreases exponentially with the size of the weights.

This asymptotically exact solution consists of plaquette-shaped filters with three positive weights and one negative weight (each of magnitude $W$) or vice-versa as shown in Fig.~\ref{fig: toric code}, which we call "odd parity filters". We can interpret this structure by considering an exact toric code ground state solution, which consists of an equal superposition of all spin-configurations $|\bm{ \sigma}\rangle$ that fulfill the ``plaquette conditions'' $\lambda_{B_p}(\bm{\sigma}) = \prod_{i \in p} \sigma^z_i = 1$ $\forall p$ \cite{kitaev2003fault}. In other words, to represent this ground state a neural-network solution has to fulfill $\Psi^{NN}(\bm{\sigma})=1$ if $\lambda_{B_p}(\bm{\sigma})=1$ $\forall p$, and $\Psi^{NN}(\bm{\sigma})=0$ else. Then, together with its activation function, an odd parity filter connected to plaquette $p$ has the effect of exponentially amplifying the amplitude $\Psi^{NN}(\bm{\sigma})$ for spin-configurations $\bm{\sigma}$ that fulfill $\lambda_{B_p}(\bm{\sigma})=1$ relative to configurations with $\lambda_{B_p}(\bm{\sigma})=-1$ (see Table \ref{table: toric code exact solution}). In total, the excited state contributions are suppressed relative to the ground state by a factor of $\exp(4W)$ for every broken plaquette condition. In the limit of $W\to \infty$, we thus arrive at the exact solution. Further details on the convergence of this solution over increasing magnitude weights can be found in the Appendix.

\begin{table}
  \label{table: toric code exact solution}
  \caption{Explanation of our neural network solution to the toric code. The two rows consider configurations of spins that obey/violate the plaquette condition. The third column shows the average post-activation value of the filter convolved with the spin configuration. The last column shows the total contribution to $\Psi^{NN}$ after all 8 orientations of the filter have been summed over.} 
  \centering
  \begin{tabular}{llll}
    \toprule
Spin config. of $\bm{\sigma}$ of single plaquette $p$ & In ground state & $\langle \textrm{ReLu}(f \cdot 
\bm{\sigma}) \rangle$ & $\exp (\sum_f \textrm{ReLu}(f \cdot 
\bm{\sigma}))$ \\
\midrule
$(4 \uparrow )$, $(2 \uparrow, 2 \downarrow)$, $(4 \downarrow )$ & \cmark & $W$ & $\exp(8W)$\\
$(3 \uparrow, 1 \downarrow)$, $(1 \uparrow, 3 \downarrow)$  & \xmark & 0.5$W$ & $\exp(4W)$\\
    \bottomrule
  \end{tabular}
\end{table}

\section{Conclusion}
\label{sec:conclusion}

We study the dynamics of neural network pruning in a novel context, the problem of finding the ground state eigenvector of a quantum many-body system. We find that the problem and the method are symbiotic; by pruning we can discover how hard certain quantum systems are to represent, and through studying the physics of the sparse subnetwork representations we can understand the mechanisms by which neural networks make approximations.   

In agreement with other pruning studies, we find that we are able to remove a large portion of neural network weights without any degradation in accuracy. Surprisingly, we find that the accuracy of models trained in isolation depends strongly on the connective structure of the weights, while the dependence on the parameter initialization is marginal. One possible explanation lies in a potentially strong connection between the sparse subnetwork connectivity and the physics of the model. Another explanation is the utilization of natural gradient descent, which can take larger steps across barren plateaus and may be able to overcome unfavorable initial parameter values.

We dig deeper into the dynamics of pruning on the transverse field Ising model, finding several fascinating features. Most saliently, at the critical point we find a first order phase transition \textit{as a function of parameters per spin} from the critical phase to a state that breaks $\mathbb{Z}_2$ spin polarization symmetry.

Finally, we show that through pruning we are able to discover a new (asymptotically exact) solution to the toric code which is simpler than human-conceptualized solutions \cite{carrasquilla2017machine, chen2025representing, deng2017machine, valenti2022correlation}, and does not require a priori architecture design principles. This demonstrates that pruning neural networks can elucidate how they are able to solve problems. We expect that pruning may uncover further exact solutions to other Hamiltonians, including those with topologically ordered ground states, those representing quantum error correction codes.

Understanding how to obtain a neural-network quantum state representation with minimal number of parameters may be relevant when it is beneficial to have a compressed model for further processing. For example, time evolution \cite{sinibaldi2023unbiasing}, quantum circuit simulation \cite{medvidovic2021classical}, and fine-tuning neural network quantum states \cite{rende2024fine} are all applications where an accurate compressed model could be advantageous. Furthermore, sparse network representations of quantum states may unlock the capability to use desired optimization strategies. Equipped with a pruned neural network that represents the initial state accurately, the fact that the architecture utilizes a small amount of parameters may allow for the use of SR in the $n<s$ regime where $n$ the number of parameters and $s$ is the number of samples. The ability to use SR may provide benefits over minSR \cite{chen2024empowering}, a variant of SR which is more efficient in the $n>s$ regime, but limited in terms of the number of samples one can use in practice. We leave the exploration of this idea to future studies.

\section*{Code and data availability}
The code, all pretrained models, and data are available on Zenodo \cite{anonymous_2025_15492729}.

\section*{Acknowledgements}
We would like to thank Bartholomew Andrews, Ao Chen, Anna Dawid, Matija Medvidović, Roger Melko and Schuyler Moss for useful discussions. The Center for Computational Quantum Physics at the Flatiron Institute is supported by the Simons Foundation. The computations in this work were, in part, run at facilities supported by the Scientific Computing Core at the Flatiron Institute. The Flatiron institute is a division of the Simons Foundation.

% \begin{ack}
% Use unnumbered first level headings for the acknowledgments. All acknowledgments

% Do {\bf not} include this section in the anonymized submission, only in the final paper. You can use the \texttt{ack} environment provided in the style file to automatically hide this section in the anonymized submission.

% \end{ack}

\bibliographystyle{plain}
\bibliography{main.bib}

%%%%%%%%%%%%%%%%%%%%%%%%%%%%%%%%%%%%%%%%%%%%%%%%%%%%%%%%%%%%

% \appendix

% \section{Technical Appendices and Supplementary Material}
% Technical appendices with additional results, figures, graphs and proofs may be submitted with the paper submission before the full submission deadline (see above), or as a separate PDF in the ZIP file below before the supplementary material deadline. There is no page limit for the technical appendices.

\newpage
\appendix

\section{Pruning procedures}\label{app:pruning}

\begin{algorithm}[h]
    \caption{Iterative Magnitude Pruning with Weight Rewinding (IMP-WR) \cite{frankle2018lottery}}\label{alg:IMP-WR}
    \begin{algorithmic}
        \STATE \textbf{Input:} A randomly initialized dense network $f(\bm{\sigma}; \theta_{\mathrm{init}} \odot m)$ with binary mask $m = \{1\}^{|\theta_{\mathrm{init}}|}$.
        \STATE \textbf{Output:} A sparse network $f(\bm{\sigma}; \theta^{\prime} \odot m^{\prime})$.
        \STATE Train the network for $j$ steps. \COMMENT{Pre-training phase}
        \STATE Save the parameters $\theta_{\mathrm{wr}}$ for rewinding.
        \FOR{each pruning iteration in $I$}
            \STATE Select $p_{\mathrm{r}}$\% of weights (among unmasked) with lowest magnitude.
            \STATE Set corresponding entries in the mask $m$ to zero.
            \STATE Reset unmasked weights to their values in $\theta_{\mathrm{wr}}$. \COMMENT{Weight rewinding}
            \STATE Train the network for $k$ steps.
        \ENDFOR
    \end{algorithmic}
\end{algorithm}

\paragraph{Comparison of IMP-WR to other pruning procedures}

\begin{figure}[h]
    \centering
    \includegraphics[width=\linewidth]{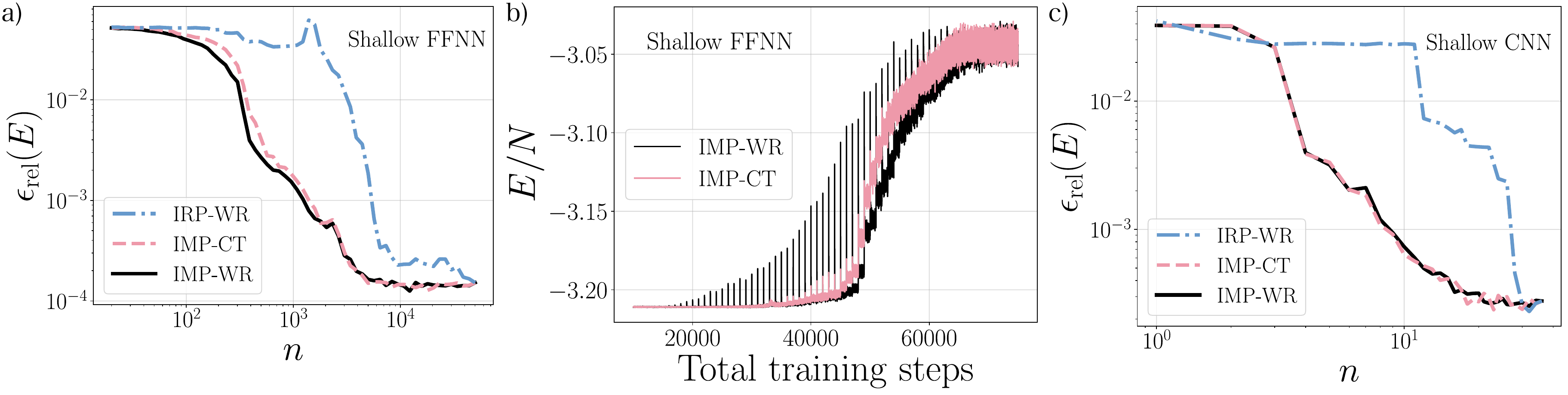}
    \caption{Comparison of iterative pruning variants: weight rewinding (IMP-WR), continued training (IMP-CT), and random pruning with weight rewinding (IRP-WR). This data corresponds to the TFIM on a $N=10\times10$ lattice at the critical point ($\kappa = \kappa_c$). We plot the relative error of these three variants starting from the same dense (a) FFNN $(d=1, w=5N)$, and (c) CNN $(d=1, n_f=4)$. In (b), we show the energy per spin over the total training steps in 65 pruning iterations from the shallow FFNN in (a).}
    \label{fig: comparison of pruning methods}
\end{figure}

In this work, we use Iterative Magnitude Pruning with Weight Rewinding (IMP-WR) as described in Algorithm~\ref{alg:IMP-WR} and Section 3.3 in the main text. We compare this baseline with two iterative pruning variants:

\begin{itemize}
    \item \textbf{IMP-CT (Continued Training):} Follows the same procedure as IMP-WR, except that after each pruning step, the remaining weights are not rewound to their initial values. Instead, training continues from their current values. This approach is often referred to as \emph{fine-tuning}.
    
    \item \textbf{IRP-WR (Iterative Random Pruning with Weight Rewinding):} Also mirrors the IMP-WR procedure, but replaces magnitude-based pruning with uniform random selection of weights to prune. This variant serves as a baseline to assess the efficacy of magnitude-based criteria.
\end{itemize}

For shallow FFNNs, we find that IMP-WR converges to a lower energy minima at each pruning iteration, compared to IMP-CT as shown in Fig. \ref{fig: comparison of pruning methods}, in agreement with previous work comparing weight rewinding and continued training \cite{renda2020comparing}. Despite the large perturbation of the network due to rewinding the weights shown in Fig. \ref{fig: comparison of pruning methods}, the network is able to recover to a lower energy with few training steps at each pruning iteration. Conversely, continued training does not induce such a large perturbation, and the network retains an energy close to the ground state, yet these minima are higher in energy than those of IMP-WR. Compared to the baseline of IRP-WR, we find that networks pruned on magnitude based criteria in IMP-WR and IMP-CT drastically outperform those pruned randomly with IRP-WR. This result further demonstrates that magnitude-based pruning criteria finds a non-trivial structure in the sparse subnetworks, that achieves significantly better than a random structure. The differences between weight rewinding and continued training are not as distinct for shallow CNNs compared to shallow FFNNs, shown in Fig. \ref{fig: comparison of pruning methods}, which is likely due to the very number of weights in the model from initialization.

\section{Scaling behavior in the variance of the energy}

\begin{figure}[h]
    \centering
    \includegraphics[width=\linewidth]{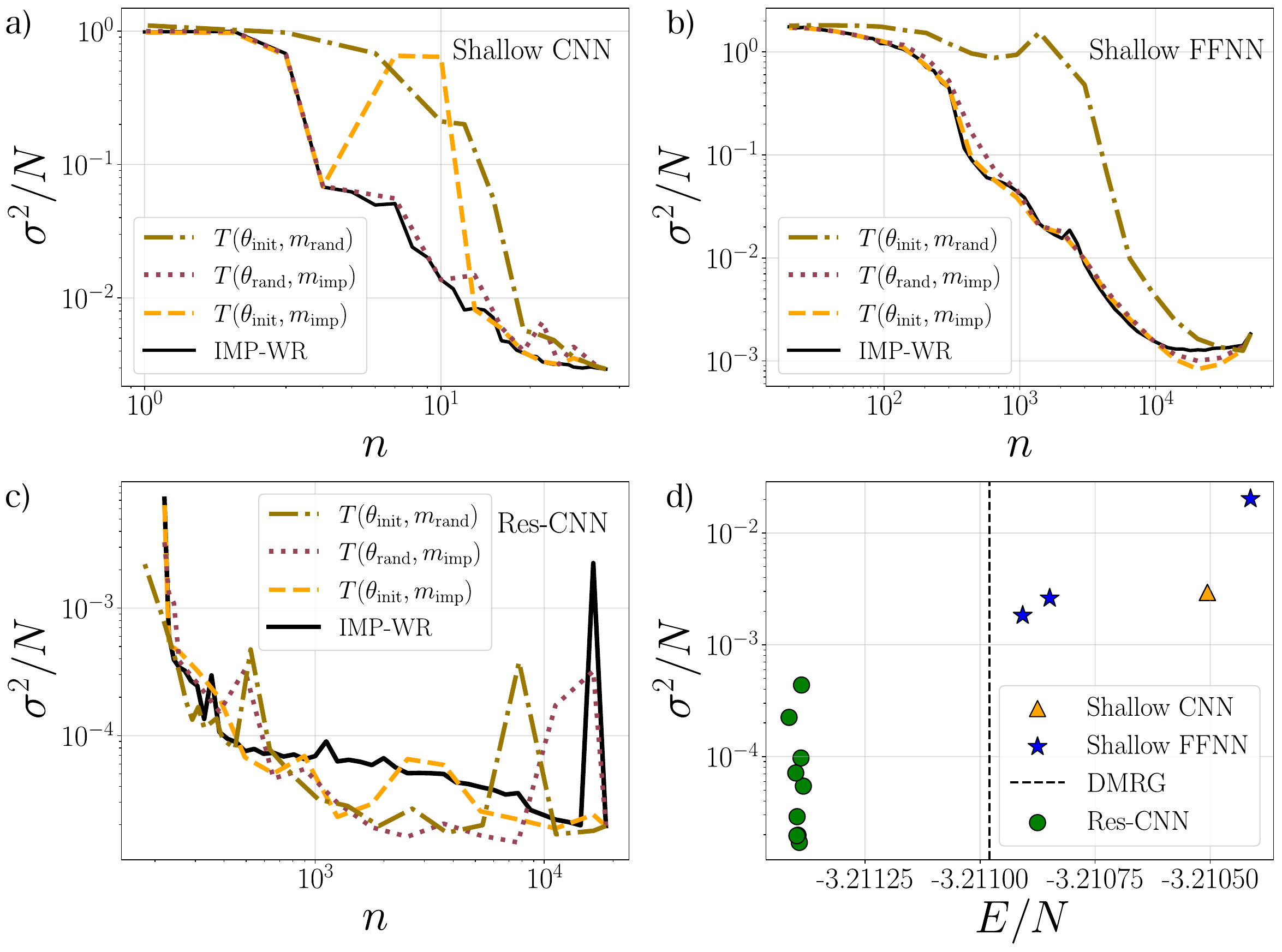}
    \caption{Variance of the energy per spin $\sigma^2/N$ over remaining parameters $n$ corresponding to the data in Fig. \ref{fig: LTH testing} in the main text, for the $N=10\times 10$ TFIM at the critical point ($\kappa = \kappa_c$). We include IMP-WR and isolated training variants from various networks: (a) shallow FFNN, (b) shallow CNN, and (c) ResCNN. In (d), we show extended data from the inset in Fig. 1 for theenergy variance per spin over the energy per spin $E/N$ associated to various dense initializations of single layer FFNNs varying in width $w=\alpha N;\alpha=\{1,5/2,5\}$, single layer CNNs varying in width by the number of filter $n_f=\{4,8,16\}$, and ResCNNs varying in depth and width. In (d), we include the energy per spin obtained from DMRG with bond dimension $\chi=3000$ as a vertical dashed line.}
    \label{fig: variances}
\end{figure}

In this section, we report the energy variance per spin for the iteratively pruned and isolated trained networks presented in the main text. We define the energy variance as
\begin{equation}
    \sigma^2 = \langle \hat{H}^2 \rangle - \langle \hat{H} \rangle^2
\end{equation}
which is estimated as the sample variance of the local energy defined in Eq. \ref{eq: local_expec}. This quantity controls the statistical uncertainty $\sigma_{E}$ in the estimated energy:
\begin{equation}
    \sigma_{E} = \sqrt{ \frac{\sigma^2}{N_{\mathrm{s}}} },
\end{equation}
where $N_s$ is the number of samples. As a result, the energy variance provides a meaningful measure of the magnitude of fluctuations in $\langle \hat{H} \rangle$ and the quality of the ground-state approximation, as it vanishes when the variational wavefunction is an exact eigenstate and governs the statistical error in energy estimates obtained via Monte Carlo sampling.

We find that the scaling behavior of the energy variance follows the same scaling behavior of the relative error of the energy. This demonstrates that our conclusions drawn from the scaling of relative error are well supported, and not simply due to fluctuations in the sampled energy estimates.

\section{Robustness over many random initializations}

\begin{table}
  \caption{Robustness of $m_{\mathrm{imp}}$ mask structure at different iterations of pruning for the $N=10\times 10$ TFIM (corresponding to panel (b) of Fig. \ref{fig: LTH testing}), and $N=18$ toric code (corresponding to panel (a) of Fig. \ref{fig: toric code}) Hamiltonians. We report the original relative energy error for $\theta_{\mathrm{init}}$ and the mean relative energy error over 5 different initializations for $\theta_{\mathrm{rand}}$, in addition to their 99.9\% confidence interval.}
  \label{table: robustness over random inits}
  \centering
  \begin{tabular}{ccccc}
    \toprule
\multirow{2}{*}{Hamiltonian} & \multirow{2}{*}{Pruning iteration} & \multicolumn{2}{c}{Initialization type} & \\
&  & $\theta_{\mathrm{init}}$ & $\theta_{\mathrm{rand}}$ & 99.9\% CI \\
\midrule
\multirow{3}{*}{TFIM} & $i=12$ & 1.54e-4 & 1.50e-4 & [1.44e-4, 1.56e-4] \\
& $i=15$ & 1.65e-4 & 1.67e-4 & [1.58e-4, 1.77e-4]\\
& $i=18$ & 1.96e-4 & 1.99e-4 & [1.50e-4, 2.19e-4]\\
\midrule
\multirow{3}{*}{TC} & $i=9$ & 2.94e-5 & 1.83e-5 & [0, 4.55e-5]\\
 & $i=12$ & 1.64e-5 &  1.52e-5 & [0, 3.45e-5]\\
  & $i=15$ & 9.71e-6 & 1.60e-5 & [0, 4.30e-5]\\
    \bottomrule
  \end{tabular}
\end{table}

To substantiate the claim of the importance of the structure in the masks, and the relative unimportance of parameter initialization, we test the parameter sensitivity in the reduced model dimension for the $T(\theta_{\mathrm{rand}}, m_{\mathrm{imp}})$ isolated training tests, starting from many random initializations. We confirm the robustness of the subnetwork structure by reporting the mean of five additional random instances of the isolated training variant at several pruning iterations and their 99.9\% confidence interval shown in Table \ref{table: robustness over random inits}. All random initializations remain close to the $\theta_{\mathrm{init}}$ initialization and are in agreement at the 99.9\% level. This confirms the robustness of the mask structure, and demonstrates that under many random initializations, a good mask is sufficient to obtain a matching error to the lottery ticket initialization.

\section{Sparse neural network solution to the toric code}

\begin{figure}[h]
    \centering
    \includegraphics[width=\linewidth]{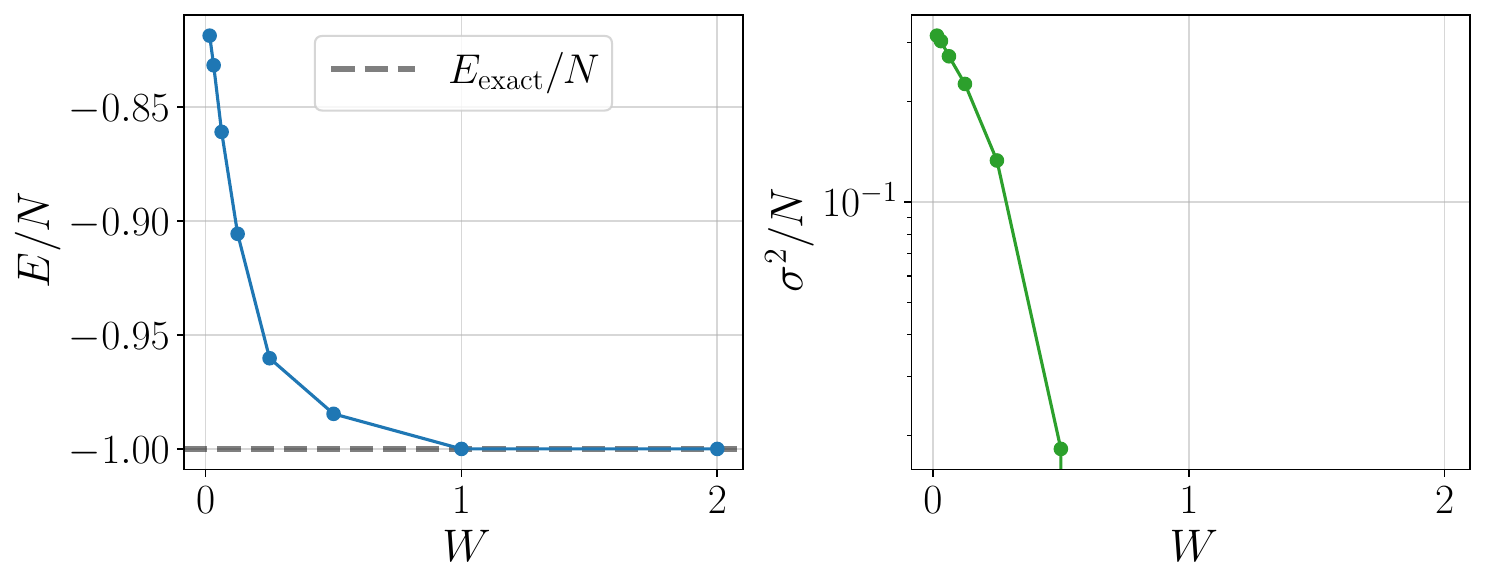}
    \caption{Exponential convergence to the ground state of the toric code, driven by the magnitude of the weights $W = \vert w_{ij} \vert$ in odd-parity filters within single hidden neurons of $d=1, w=4N$ FFNN. In (a), we plot $E/N$, the energy per spin from \emph{untrained} sparse initialized networks with odd-parity single hidden neuron filters. In (b), we plot the energy variance, which abruptly goes to zero, signifying an exact solution to the toric code.}
    \label{fig: tc exp convergence}
\end{figure}

In this section, we further detail the sparse FFNN solution to the toric code. In Section 4.4 of the main text, we provide a single hidden neuron analysis for a single plaquette spin configuration. We show that odd-parity filters exponentially amplify the wavefunction coefficients of ground state configurations by the magnitude of positive and negative weights in odd-parity filters.

In Fig. \ref{fig: tc exp convergence}, we demonstrate this exponential convergence by computing the energy from a sparse initialized networks with odd-parity filter structure. When the magnitude of the weights increase, we find that the energy converges to the exact ground state energy, with zero variance.

\section{Neural network architectures}\label{app:architectures}

\begin{figure}[h]
    \centering
    \includegraphics[width=\linewidth]{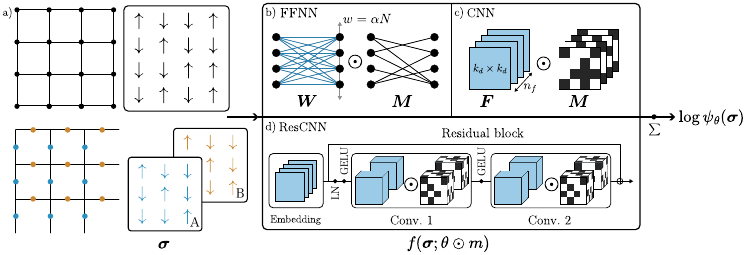}
    \caption{Schematic of neural network architectures from (a) input spin configurations $\bm{\sigma}$, to (b,c,d) neural network ansatz $f(\bm{\sigma}; \theta \odot m)$, which outputs wavefunction coefficients $\log \psi_{\theta}(\bm{\sigma})$. In (a), we show the input configurations of the (top) TFIM and (bottom) A and B sublattices of toric code, which are all flattened upon input to FFNNs. We detail the masking procedure in (a) single layer FFNN, (b) single layer CNN, and (c) ResCNN with a single residual block. Each residual block in (c) consists of a layer normalization (LN), GELU activation, convolutional layer (Conv. 1), GELU activation, convolutional layer (Conv. 2), and residual connection. The outputs of the last layer are summed, to output the logarithm of the wave function coefficient.}
    \label{fig: network architectures}
\end{figure}

\paragraph{Feed-forward neural networks}
Throughout this work, we use a shallow feed-forward neural network (FFNN) architecture of constant depth $d=1$, varying in width $w = \alpha N$, where $\alpha$ is a scaling factor of the input dimension. We choose to scale the size of the networks with the system-size, such that the input dimension $N$ denotes the number of spins, and the latent dimension is scaled by $\alpha \in \mathbb{R}$. Therefore, the set of parameters $\theta$ are simply defined by a synaptic weight matrix $\bm{W} \in \mathbb{R}^{N \times \alpha N}$. We initialize $\bm{W}$ from a normal distribution with a standard deviation $\sigma=0.1$. The network has bias terms $\bm{b}$, pre-activations $\bm{h}$, and post-activations $\tilde{\bm{\sigma}}$. The forward propagation dynamics of the neural network are given by 
\begin{equation}
    \tilde{\bm{\sigma}} = \phi(\bm{h}), \quad \bm{h} = \bm{W}\bm{\sigma} + \bm{b},
    \label{eq:ffnn_propogation}
\end{equation}
where $\phi: \mathbb{R} \rightarrow \mathbb{R}$ is a non-linear activation function, and the input to the network is $\bm{\sigma} \in \mathbb{R}^N$. We use the Rectified Linear Unit (ReLU) activation function throughout this work. For the simplicity of interpreting weights without the effect of bias, we choose to remove the bias terms, effectively setting $\bm{b}=0$ in Eq. \ref{eq:ffnn_propogation}. The removal of bias terms produced no noticeable differences in network performance.
The post activations are then summed to produce the logarithm of the wave function coefficients, therefore giving the exponential ansatz defined by 
\begin{equation}
    \psi(\bm{\sigma}) = \exp\left[\sum_{i=1}^{\alpha N} \tilde{\bm{\sigma}}_i\right].
    \label{eq:ffnn_exp_ansatz}
\end{equation}
The total number of parameters in the network at dense initialization is $n_{\mathrm{init}} = \alpha N^2$. Thus, the number of parameters in this shallow FFNN architecture scales as $\mathcal{O}(N^2)$, which is quadratic in the number of spins.

\paragraph{Shallow convolutional neural networks}
We use a shallow convolutional neural-network (CNN) architecture of constant $d=1$ depth, varying in width by $n_f$, the number of convolutional kernels. The forward propagation dynamics of the neural network are given by 
\begin{equation}
    \tilde{\bm{\sigma}} = \phi(\bm{h}), \quad \bm{h} = \bm{F} *\bm{\sigma} + \bm{b},
    \label{eq:cnn_propogation}
\end{equation}
where $\bm{F} * \bm{\sigma}$ denotes the convolution operation of $n_f$ kernels $f \in \bm{F}$. The kernels are convolved with VALID padding \cite{flax2020github} on the two-dimensional input $\bm{\sigma}$. The kernels $f \in \mathbb{R}^{k_d \times k_d}$ have dimension $k_d = 3$, which we initialize with a truncated normal distribution (LeCun normal) \cite{klambauer2017self}. Similar to the FFNN architecture, we set the bias terms to zero. The activation $\phi$ we use is a Gaussian Error Linear Unit (GELU). The post activations are summed in a similar manner to Eq. \ref{eq:ffnn_exp_ansatz} giving
\begin{equation}
    \psi(\bm{\sigma}) = \exp\left[\sum_{i=1}^{N} \tilde{\bm{\sigma}}_i\right].
    \label{eq:cnn_exp_ansatz}
\end{equation}
The total number of parameters of a CNN at dense initialization is therefore $n_{\mathrm{init}} = n_f k_d^2$, which scales as a constant with the system size $N$, due to the nature of weight sharing in the convolution operation.

\paragraph{Residual convolutional neural networks}
We use a residual layer convolutional neural network (ResCNN) introduced in \cite{chen2024empowering, chen2025convolutional} that vary by (width) the number convolutional kernels $n_f$, and (depth) the number of residual blocks $n_b$. The ResCNN architecture consists of an initial convolutional embedding layer, followed by pre-activation residual blocks with two convolutional layers per block. We detail the structure of the ResCNN in Fig. \ref{fig: network architectures}. The embedding layer of the CNN produces the input to the residual blocks
\begin{equation}
    \bm{h}^{1} = \bm{F}_{\mathrm{embed}} *\bm{\sigma},
    \label{eq:embedding}
\end{equation}
where $\bm{F}_{\mathrm{embed}} \in \mathbb{R}^{n_f \times k_d \times k_d}$ is the embedding convolution. We use no biases for the ResCNN architecture, similar to the FFNN and CNN architectures. The forward propagation dynamics of the residual blocks are given by
\begin{equation}
    \bm{h}^{i+1} = \left( \bm{F}_2 * \phi \left\{ \bm{F}_1 * \mathrm{LN}\left[\phi(\bm{h}^i)\right] \right\} \right) + \bm{h}^i, \quad i=1, \dots, n_b.
    \label{eq:res_block}
\end{equation}
Here, LN denotes a layer normalization, $\phi$ is the GELU activation, and $\bm{F}_1, \bm{F}_2 \in \mathbb{R}^{n_f \times n_f \times k_d \times k_d}$ denote two convolutional layers (Conv. 1 and Conv.2) in each residual block. We use a custom layer normalization that computes the mean and variance only on non-zero inputs, due to the sparsity induced from pruning. We found that masking zero inputs in the layer normalization produced more stable training dynamics. All convolutional layers are initialized from a truncated normal distribution (LeCun normal) \cite{klambauer2017self}. Upon obtaining $\bm{h}^{n_b}$ from the last residual block, we apply a final layer normalization to obtain $\tilde{\bm{\sigma}} = \mathrm{LN}\left[ \bm{h}^{n_b} \right]$, and sum the outputs to produce the wave function coefficients using Eq. \ref{eq:cnn_exp_ansatz}.

The total number of parameters in a ResCNN is defined by $n_{\mathrm{init}} = n_{\mathrm{embed}} + n_b n_f^2 k_d^2$, where the $n_{\mathrm{embed}} = n_f k_d^2$ denotes the number of parameters in the embedding layer. We do not prune the embedding layer, and prune the residual blocks in a global unstructured fashion.

\section{Sampling and optimization procedures \label{app:optimization}}

\paragraph{Monte Carlo sampling}

To estimate expectation values of observables such as the energy or its gradient, we sample configurations $\bm{\sigma}$ from the probability distribution $|\psi_\theta(\bm{\sigma})|^2$ using the Metropolis-Hastings algorithm~\cite{hastings1970monte}. New configurations $\bm{\sigma}'$ are proposed via local update rules and accepted with probability
\begin{equation}
    P(\bm{\sigma} \rightarrow \bm{\sigma}') = \min\left(1, \frac{|\psi_\theta(\bm{\sigma}')|^2}{|\psi_\theta(\bm{\sigma})|^2} \right).
\end{equation}
Provided the proposal distribution is symmetric, this procedure yields samples from the desired distribution after thermalization.

The choice of proposal distribution depends on the physical model. For the transverse field Ising model, we use single-spin flips. For the toric code, where single flips have vanishing amplitude in the ground state sector, we instead propose updates from two rules with equal probability: single-spin flips and \emph{plaquette flips}, where all four spins on a plaquette are flipped simultaneously. This ensures efficient exploration of relevant ground state configurations.

\paragraph{Efficient computation of expectation values}

Expectation values of an operator $\hat{O}$ are then estimated by
\begin{equation}
    \langle \hat{O} \rangle \approx \frac{1}{N_s} \sum_{i=1}^{N_s} O_{\mathrm{loc}}(\bm{\sigma}_i),
\end{equation}
as the sample mean over $N_s$ configurations drawn from $\vert \Psi_{\theta}^{NN}(\bm{\sigma})\vert^2$. Here,
\begin{equation}
    O_{\mathrm{loc}}(\bm{\sigma}) = \frac{\langle \bm{\sigma} | \hat{O} | \Psi_{\theta}^{NN} \rangle}{\langle \bm{\sigma} | \Psi_{\theta}^{NN} \rangle},
    \label{eq: local_expec}
\end{equation}
is a local expectation value, $\vert \Psi_{\theta}^{NN} \rangle$ is the neural network quantum state, and $\vert \bm{\sigma}\rangle$ denotes a basis configuration. 

\paragraph{Stochastic reconfiguration}

We optimize neural network quantum states using stochastic reconfiguration (SR)~\cite{sorella1998green}, a technique equivalent to natural gradient descent in the variational parameter space. This method accounts for the geometry of the Hilbert space by introducing a metric tensor, ensuring more stable and efficient optimization trajectories compared to standard gradient descent.

Given a wavefunction $\vert \Psi_\theta \rangle$ parametrized by $\theta$, the SR update rule is
\begin{equation}
    \theta \rightarrow \theta - \eta \sum_{\theta'} S^{-1}_{\theta \theta'} \, \frac{\partial \langle \hat{H} \rangle_\theta}{\partial \theta'} \,,
\end{equation}
where $\eta$ is the learning rate (LR), $\langle \hat{H} \rangle_\theta = \langle \Psi_\theta \vert \hat{H} \vert \Psi_\theta \rangle / \langle \Psi_\theta \vert \Psi_\theta \rangle$ is the variational energy, and $S_{\theta \theta'}$ is the quantum Fisher information matrix (or S-matrix), defined as
\begin{equation}
    S_{\theta \theta'} = \langle O_\theta^* O_{\theta'} \rangle - \langle O_\theta^* \rangle \langle O_{\theta'} \rangle\,,
    \label{eq: S matrix}
\end{equation}
with $O_\theta(\bm{\sigma}) = \partial \log \Psi_\theta(\bm{\sigma})/\partial \theta$ denoting the log-derivatives of the wavefunction and expectations taken over the distribution of $|\Psi_\theta(\bm{\sigma})|^2$. The gradient of the energy with respect to the parameters is given by:
\begin{equation}
    \frac{\partial \langle \hat{H} \rangle_\theta}{\partial \theta} = 2 \, \text{Re} \left[ \langle O_\theta^* E_{\mathrm{loc}} \rangle - \langle O_\theta^* \rangle \langle E_{\mathrm{loc}} \rangle \right],
\end{equation}
where the local energy is defined by Eq. \ref{eq: local_expec}. To ensure that the inverse of the S-matrix in Eq. \ref{eq: S matrix} is well defined, we use an explicit regularization $S = S + \lambda I$, where $\lambda$ is called the diagonal shift (DS). We keep the learning rate and diagonal shift constant throughout this work, and report the values we use in Section \ref{app: reproducability}.

The SR method can also be derived as an approximation to imaginary-time evolution, $\vert \Psi_{t+1} \rangle \approx \exp(-\eta \hat{H}) \vert \Psi_t \rangle$, where $\eta$ plays the role of an imaginary-time step. For further details on this connection, see \cite{sorella1998green, park2020geometry, medvidovic2024neural}.

\section{Resources for reproducability}\label{app: reproducability}

\begin{table}[ht]
\centering
\caption{Symbols, acronyms, and hyperparameters used throughout this work.}
\begin{tabular}{cll}
\toprule
Symbol & Description & Applies to \\
\midrule
$\bm{\sigma}$ & Spin configuration & - \\
$N$ & Number of spins & - \\
$E$ & Variational energy & - \\
$\Delta$ & Absolute error & - \\
$\epsilon_{\mathrm{rel}}$ & Relative error & - \\
$\sigma^2$ & Energy variance & - \\
$\theta$ & Neural network parameters & - \\
$w$ & Individual neural network weight & - \\
$W$ & Magnitude of a neural network weight & - \\
$\Psi_{\theta}^{NN}$ & Neural network quantum state & - \\
$F$ & Monte Carlo fidelity & - \\
$M^{\mathrm{tot}}_{x/z}$ & Magnetization in the $x$ or $z$ directions & - \\
$T$ & A sparse neural network trained in isolation & - \\
$\theta_{\mathrm{init}}$ & Initialized parameters from IMP-WR & - \\
$\theta_{\mathrm{rand}}$ & Randomly initialized parameters & - \\
$m_{\mathrm{imp}}$ & Sparse mask obtained from from IMP-WR & - \\
$m_{\mathrm{rand}}$ & Random sparse mask & - \\
$n$ & Number of remaining parameters & - \\
$n_{\mathrm{init}}$ & Number of initial parameters & - \\
$\rho$ & Number of parameters per spin & - \\
$\sigma^{x/z}$ & Pauli matrices & - \\
$\kappa$ & Transverse field strength & - \\
TFIM & Transverse field Ising model & - \\
TC & Toric code & - \\
LTH & Lottery ticket hypothesis & - \\
DMRG & Density matrix renormalization group & - \\
FFNN & Feed forward neural-network & - \\
CNN & Convolutional neural-network & - \\
ResCNN & Residual convolutional neural-network & - \\
IMP-WR & Iterative magnitude pruning with weight rewinding & - \\
IMP-CT & Iterative magnitude pruning with continued training & - \\
IRP-WR & Iterative random pruning with weight rewinding & - \\
LEP & Low-error plateau & - \\
APL & Approximate power law & - \\
HEP & High-error plateau & - \\
$\bm{W}$ & Weight matrix & FFNN \\
$\bm{b}$ & Bias vector & FFNN \\
$w$ & Network width & FFNN \\
$\alpha$ & Width scaling factor & FFNN\\
$d$ & Network depth & FFNN, CNN \\
$f$ & Individual convolutional kernel & CNN, ResCNN \\
$\bm{F}$ & Convolutional kernel & CNN, ResCNN \\
$n_f$ & Number of kernels (width) & CNN, ResCNN \\
$k_d$ & Kernel dimension & CNN, ResCNN \\
$n_b$ & Number of residual blocks (depth) & ResCNN \\
$n_{\mathrm{embed}}$ & Number of parameters in embedding layer & ResCNN \\
LN & Layer normalization & ResCNN \\
$j$ & Number of pre-training steps & All \\
$k$ & Number of training steps in each pruning iteration & All \\
$p_r$ & Pruning ratio & All \\
$I$ & Total number of pruning iterations & All \\
$n_{\mathrm{init}}$ & Number of parameters at dense initialization & All \\
$N_s$ & Number of samples per optimization step & All \\
$\eta$ & Learning rate (LR) & All \\
$\lambda$ & Diagonal shift (DS) in stochastic reconfiguration & All \\
\bottomrule
\end{tabular}
\label{tab:hyperparams}
\end{table}

\paragraph{Hyperparameters and training settings}
 All symbols, acronyms, and hyperparameters used in this work are reported in Table \ref{tab:hyperparams}. We report the all hyperparameters and training settings used across all experiments in Tables \ref{table:fig1 hyperparams}, \ref{table:fig2 hyperparams}, \ref{table:fig3 hyperparams}, and \ref{table:fig4 hyperparams} to ensure reproducibility and facilitate comparison in future work.

\begin{table}[h]
\caption{Hyperparameters used in Fig. \ref{fig: LTH testing} in the main text, for each architecture in panels (a) CNN, (b) FFNN, and (c) ResCNN. Definitions of all symbols are provided in Table~\ref{tab:hyperparams} in the appendix.}
\label{table:fig1 hyperparams}
\begin{center}
\begin{small}
\begin{sc}
\begin{tabular}{lccccccccccr}
\toprule
Network & $w/n_f$ & $d/n_b$ & $k_d$ & $j$ & $k$ & $p_r$ & $I$ & $n_{\mathrm{init}}$ & $N_{\mathrm{s}}$ & $\eta$ & $\lambda$\\
\midrule
a) CNN & $n_f=4$ & $d=1$ & 3 & 1e4 & 1e3 & 0.05 & 31 & 36 & 1024 & 8e-3 & 1e-3\\
b) FFNN & $w= 5N$ & $d=1$ & -- & 1e4 & 1e3 & 0.12 & 65 & 50000 & 1024 & 8e-3 & 1e-4\\
c) ResCNN & $n_f=16$ & $n_b=4$ & 3 & 1e4 & 1e3 & 0.12 & 43 & 18576 & 1024 & 8e-3 & 1e-3\\
\bottomrule
\end{tabular}
\end{sc}
\end{small}
\end{center}
\end{table}

\begin{table}[h]
\caption{Hyperparameters for Fig. \ref{fig: regions of scaling} in the main text, panels (a,b,c). Definitions of all symbols are provided in Table~\ref{tab:hyperparams} in the appendix.}
\label{table:fig2 hyperparams}
\begin{center}
\begin{small}
\begin{sc}
\begin{tabular}{lccccccccccr}
\toprule
Panel & $\kappa$ & $w$ & $d$ & $j$ & $k$ & $p_r$ & $I$ & $n_{\mathrm{init}}$ & $N_{\mathrm{s}}$ & $\eta$ & $\lambda$\\
\midrule
(a) & 3.04438 & $N$ & 1 & 1e4 & 1e3 & 0.12 & 54 & 10000 & 1024 & 8e-3 & 1e-4\\
(a) & 3.04438 & $2.5N$ & 1 & 1e4 & 1e3 & 0.12 & 60 & 25000 & 1024 & 8e-3 & 1e-4\\
(a) & 3.04438 & $5N$ & 1 & 1e4 & 1e3 & 0.12 & 65 & 50000 & 1024 & 8e-3 & 1e-4\\
(b) & $\{0.1, 1, 2\}$ & $5N$ & 1 & 1e4 & 1e3 & 0.12 & 65 & 50000 & 1024 & 8e-3 & 1e-4\\
(c) & $\{4, 5, 6 \}$ & $5N$ & 1 & 1e4 & 1e3 & 0.12 & 65 & 50000 & 1024 & 8e-3 & 1e-4\\
\bottomrule
\end{tabular}
\end{sc}
\end{small}
\end{center}
\end{table}

\begin{table}[h]
\caption{Hyperparameters for Fig. \ref{fig: phase transition} in the main text, panel (a). Definitions of all symbols are provided in Table~\ref{tab:hyperparams} in the appendix.}
\label{table:fig3 hyperparams}
\begin{center}
\begin{small}
\begin{sc}
\begin{tabular}{lcccccccccr}
\toprule
System size $N$ & $w$ & $d$ & $j$ & $k$ & $p_r$ & $I$ & $n_{\mathrm{init}}$ & $N_{\mathrm{s}}$ & $\eta$ & $\lambda$\\
\midrule
$4\times 4$ & $8N$ & 1 & 1e4 & 1e3 & 0.12 & 51 & 2048 & 1024 & 8e-3 & 1e-4\\
$5\times 5$ & $8N$ & 1 & 1e4 & 1e3 & 0.12 & 58 & 5000 & 1024 & 8e-3 & 1e-4\\
$6\times 6$ & $8N$ & 1 & 1e4 & 1e3 & 0.12 & 64 & 10368 & 1024 & 8e-3 & 1e-4\\
$7\times 7$ & $8N$ & 1 & 1e4 & 1e3 & 0.12 & 69 & 19208 & 1024 & 8e-3 & 1e-4\\
$8\times 8$ & $8N$ & 1 & 1e4 & 1e3 & 0.12 & 65 & 32768 & 1024 & 8e-3 & 1e-4\\
$9\times 9$ & $8N$ & 1 & 1e4 & 1e3 & 0.12 & 74 & 52488 & 1024 & 8e-3 & 1e-4\\
$10\times 10$ & $8N$ & 1 & 1e4 & 1e3 & 0.12 & 74 & 80000 & 1024 & 8e-3 & 1e-4\\
\bottomrule
\end{tabular}
\end{sc}
\end{small}
\end{center}
\end{table}

\begin{table}[h]
\caption{Hyperparameters for Fig. \ref{fig: toric code} in the main text, panels (a,b). Definitions of all symbols are provided in Table~\ref{tab:hyperparams} in the appendix.}
\label{table:fig4 hyperparams}
\begin{center}
\begin{small}
\begin{sc}
\begin{tabular}{lcccccccccr}
\toprule
Panel(s) & $w$ & $d$ & $j$ & $k$ & $p_r$ & $I$ & $n_{\mathrm{init}}$ & $N_{\mathrm{s}}$ & $\eta$ & $\lambda$\\
\midrule
(a,b) & $4N$ & 1 & 1e4 & 1e3 & 0.12 & 47 & 1296 & 1024 & 8e-3 & 1e-3\\
(b) & $8N$ & 1 & 1e4 & 1e3 & 0.12 & 53 & 2592 & 1024 & 8e-3 & 1e-3\\
(b) & $16N$ & 1 & 1e4 & 1e3 & 0.12 & 58 & 5184 & 1024 & 8e-3 & 1e-3\\
(b) & $32N$ & 1 & 1e4 & 1e3 & 0.12 & 64 & 10368 & 1024 & 8e-3 & 1e-3\\
\bottomrule
\end{tabular}
\end{sc}
\end{small}
\end{center}
\end{table}

\paragraph{Implementation details for density matrix renormalization group}\label{app:dmrg_details}

To benchmark the quality of expected observables calculated with neural network wave functions, we used DMRG to find the ground state wave function for the TFIM. DMRG optimizes an MPS to find the lowest eigenvector of the Hamiltonian $H_{\mathrm{TFIM}}$, and the corresponding energy eigenvalue. We implemented DMRG with the ITensor julia package \cite{itensor}. The hyperparameters we used are reported in Table \ref{table: DMRG}.

\begin{table}[h]
\caption{DMRG parameters used for calculating reference observable quantities in the two-dimensional TFIM. Each sweep is bound to a maximum bond dimension $\chi_{\mathrm{max}}$, which increases at each sweep. The cutoff defines the truncation error in the MPS, which is constant for each sweep.}
\label{table: DMRG}
\vskip 0.15in
\begin{center}
\begin{small}
\begin{sc}
\begin{tabular}{lcr}
\toprule
Sweeps & Maximum bond dimension $\chi_{\mathrm{max}}$ & Cut off \\
\midrule
$12$ & $10-3000$ & 1e-15 \\
\bottomrule
\end{tabular}
\end{sc}
\end{small}
\end{center}
\vskip -0.1in
\end{table}

\paragraph{Computational resources for reproducibility}

\begin{table}[h]
\caption{Execution times (in GPU hrs) rounded up to the nearest integer for iterative pruning (IP) variants, and isolated training (IT) variants for different network architectures varying in dimension, across different system sizes of the TFIM, and toric code Hamiltonians. The execution times for isolated training variants correspond to training a single network.}
\label{table: execution times}
\vskip 0.15in
\begin{center}
\begin{small}
\begin{sc}
\begin{tabular}{lccccr}
\toprule
Hamiltonian & System size & Network & Network dim. & IP (GPU hrs) & IT (GPU hrs)\\
\midrule
TFIM & $N=4\times4$ & FFNN & $d=1, w=8N$ & 4 & 1\\
TFIM & $N=5\times5$ & FFNN & $d=1, w=8N$ & 5 & 1\\
TFIM & $N=6\times6$ & FFNN & $d=1, w=8N$ & 6 & 1\\
TFIM & $N=7\times7$ & FFNN & $d=1, w=8N$ & 10 & 2\\
TFIM & $N=8\times8$ & FFNN & $d=1, w=8N$ & 13 & 2\\
TFIM & $N=9\times9$ & FFNN & $d=1, w=8N$ & 16 & 2\\
TFIM & $N=10\times10$ & FFNN & $d=1, w=N$ & 20 & 3\\
TFIM & $N=10\times10$ & FFNN & $d=1, w=2.5N$ & 21 & 3\\
TFIM & $N=10\times10$ & FFNN & $d=1, w=5N$ & 22 & 3 \\
TFIM & $N=10\times10$ & FFNN & $d=1, w=8N$ & 23 & 3\\
TFIM & $N=10\times10$ & CNN & $d=1, n_f=4$ & 12 & 3\\
TFIM & $N=10\times10$ & ResCNN & $n_b=4, n_f=16$ & 28 & 5\\
Toric code & $N=18$ & FFNN & $d=1, w=4N$ & 1 & 1\\
Toric code & $N=18$ & FFNN & $d=1, w=8N$ & 1 & 1\\
Toric code & $N=18$ & FFNN & $d=1, w=16N$ & 1 & 1\\
Toric code & $N=18$ & FFNN & $d=1, w=32N$ & 1 & 1\\
\bottomrule
\end{tabular}
\end{sc}
\end{small}
\end{center}
\vskip -0.1in
\end{table}

All neural network training was performed on up to 30 NVIDIA A100 graphical processing units (GPUs), with 80 gigabytes (GB) of memory each, housed on an internal cluster. Iterative pruning and training as outlined in Algorithm \ref{alg:IMP-WR} was performed in serial on a single GPU, while isolated sparse training variants were carried out on multiple GPUs in parallel. The time of execution for each experiment is outline in Table \ref{table: execution times} We used an approximate total of 3200 GPU hours in this work, including 2200 GPU hours during testing, and 1000 GPU hours for final experiments. The DMRG algorithm was run on a single CPU with 80GB of memory, for a total of 30 hours.

\end{document}